\documentclass[9pt,aps,prl,footinbib,twocolumn]{revtex4-1}

\usepackage{graphicx}
\usepackage{amsmath}
\usepackage{multirow}
\usepackage{fancyvrb}

\usepackage{soul}
\usepackage{xcolor}

\begin{document}

\title{Accurate Detection of Arbitrary Photon Statistics}
\author{Josef Hlou\v{s}ek}
\affiliation{Department of Optics, Palack\'{y} University, 17. listopadu 12, 77146 Olomouc, Czechia}
\author{Michal Dudka}
\affiliation{Department of Optics, Palack\'{y} University, 17. listopadu 12, 77146 Olomouc, Czechia}
\author{Ivo Straka}
\affiliation{Department of Optics, Palack\'{y} University, 17. listopadu 12, 77146 Olomouc, Czechia}
\author{Miroslav Je\v{z}ek}
\email{jezek@optics.upol.cz}
\affiliation{Department of Optics, Palack\'{y} University, 17. listopadu 12, 77146 Olomouc, Czechia}

\begin{abstract}
We report a measurement workflow free of systematic errors consisting of a reconfigurable photon-number-resolving detector, custom electronic circuitry, and faithful data-processing algorithm. We achieve unprecedentedly accurate measurement of various photon-number distributions going beyond the number of detection channels with average fidelity 0.998, where the error is contributed primarily by the sources themselves. Mean numbers of photons cover values up to 20 and faithful autocorrelation measurements range from $g^{(2)} = 6\times10^{-3}$ to 2. We successfully detect chaotic, classical, non-classical, non-Gaussian, and negative-Wigner-function light. Our results open new paths for optical technologies by providing full access to the photon-number information without the necessity of detector tomography.
\end{abstract}


\maketitle


The probability distribution of the number of photons in an optical mode carries a great deal of information about physical processes that generate or transform the optical signal. Along with modal structure and coherence, the statistics provides full description of light.
Precise characterization of photon statistics is a crucial requirement for many applications in the field of photonic quantum technology \cite{OBrien2009} such as quantum metrology \cite{OBrien2016,Pryde2017}, non-classical light preparation \cite{Silberhorn2016,Straka2018}, quantum secure communication \cite{Shields2018}, and photonic quantum simulations \cite{Walmsley2016,Cottet2017}. Measurement of statistical properties and non-classical feautures of light also represents enabling technology for many emerging biomedical imaging and particle-tracking techniques \cite{molecules2,MPixelCamera,Hell2017}. Statistical correlations are routinely applied to quantify the non-classicality of light \cite{Mandel1977,Saleh2016}. Obtaining photon statistics requires repeated measurements using a photon-number-resolving detector (PNRD). The important parameters of PNRDs are dynamic range, speed and accuracy.

The main result of our work is a photon-statistics retrieval method based on expectation-maximization-entropy and implemented in a PNRD design that is virtually free of systematic errors. Our results show unprecedented accuracy across dozens of tested optical signals ranging from a highly sub-Poissonian single-photon state to super-Poissonian thermal light with non-negligible multi-photon content up to $n=30$. The accuracy is achieved despite leaving all systematic errors uncorrected and operating with raw data. The proposed method also provides faithful $g^{(2)}$ values \footnote{{The autocorrelation $g^{(2)}$ stems from the second-order optical coherence evaluated as a cross-correlation of intensity $I$ at zero delay, $g^{(2)}(0)=\langle I(t)^2 \rangle / \langle I(t) \rangle^2$. For the measured signals, we calculate its value from the obtained distribution of the number of photons $n$, $g^{(2)}=\langle n(n-1)\rangle/\langle n\rangle^2$.}} for states, where the commonly used Hanbury Brown--Twiss measurement would fail due to high multiphoton content \cite{Bjork2018}.

\begin{figure}[t]
\centering
\includegraphics[width=0.95 \columnwidth]{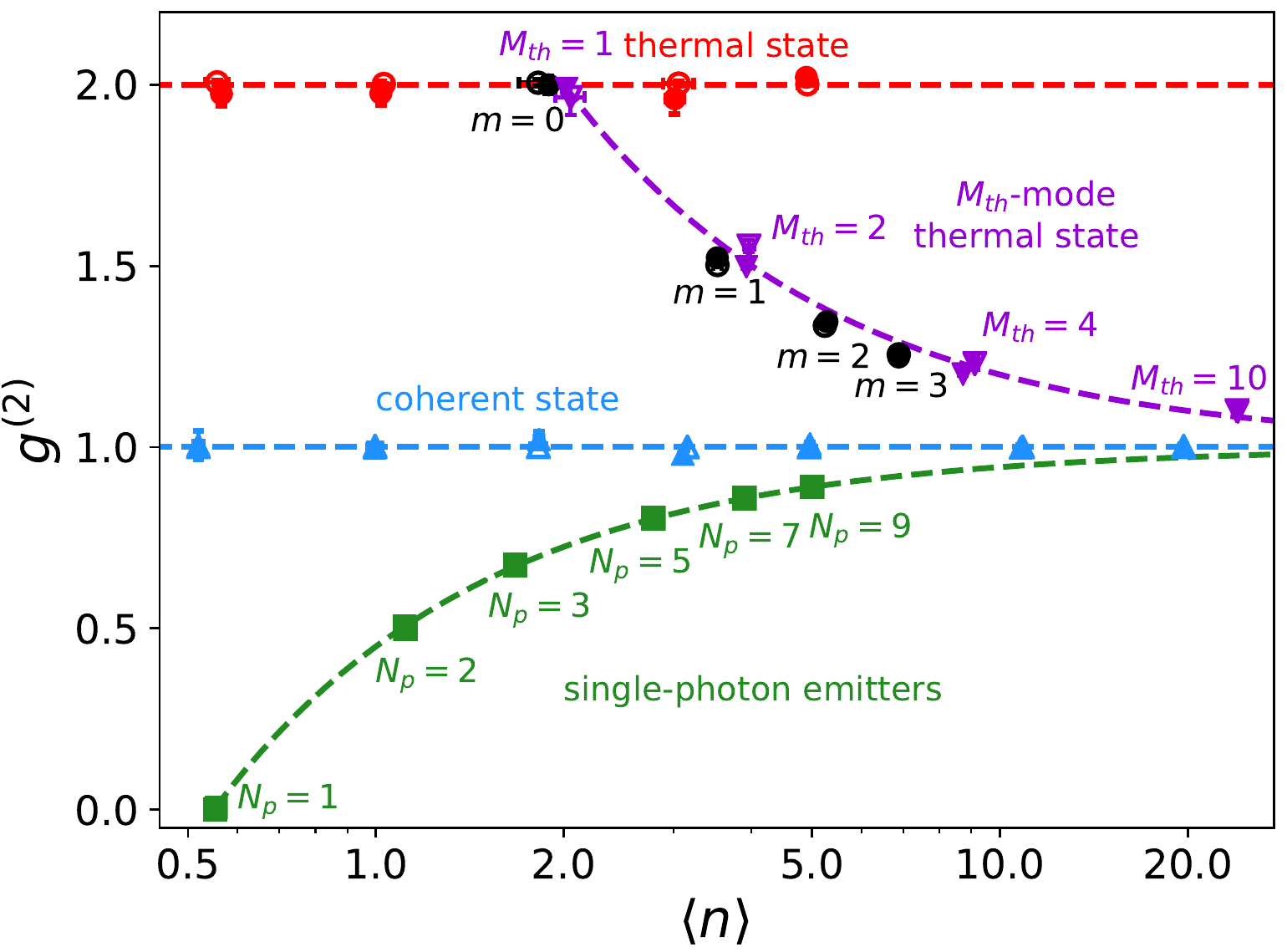}
\caption{The autocorrelation $g^{(2)}$ evaluated from the \textit{measured photon statistics} (solid marker) and the corresponding ideal statistics (empty marker) of various optical signals with mean photon number $\langle n\rangle$ \cite{Note1}. Shown are: coherent states with $g^{(2)}=1$ (blue triangle up), thermal states (also termed chaotic light) with $g^{(2)}=2$ (red circle), $M_{th}$-mode thermal states with $M_{th}=1,2,4,10$ (violet triangle down), and $m$-photon-subtracted termal states for $m=0,1,2,3$ (black circle). The cases of $M_{th}=1$ and $m=0$ coincide with thermal state. Furthermore, the emission from a cluster of $N_p$ single-photon emitters is shown for $N_p=1\ldots9$ with $g^{(2)}=1-1/N_p$.}
\label{fig_g2_vs_mean}
\end{figure}

We demonstrate the accuracy of the reported PNRD by performing photon statistics measurement for many different states of light, from which 25 states are shown in Fig.~\ref{fig_g2_vs_mean}, covering various mean photon numbers and $g^{(2)}$ values. Furthermore, the reconfigurability of the presented PNRD also allows for direct measurement of correlation functions and nonclassicality witnesses \cite{Walmsley2017,Straka2018}.

\begin{figure*}[t]
\centering
\includegraphics[width=1.0\textwidth]{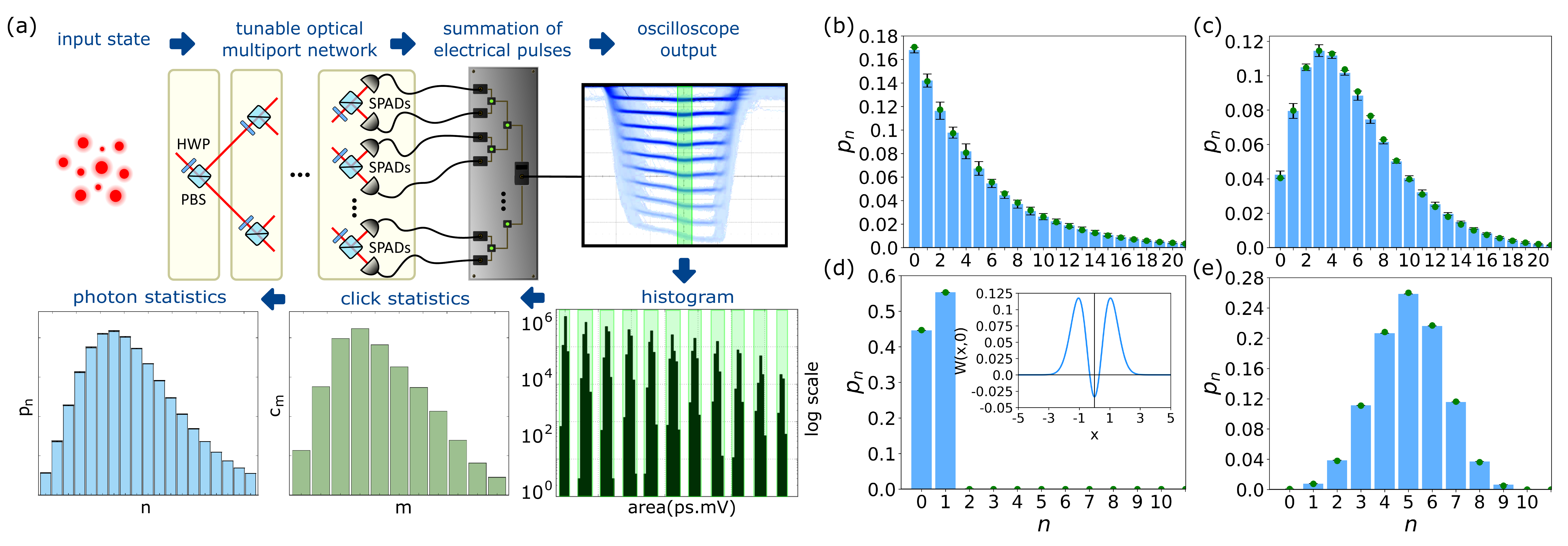}
\caption{(a) Experimental setup of the PNRD based on a discrete optical network with full reconfigurability and continuous tunability of splitting ratios; pulse-height spectrum of the analog output of the detector; and scheme of photon statistics retrieval. Measured (blue bars) and the corresponding theoretical photon statistics (green dots) for (b) thermal state with $\langle n\rangle=4.93(4)$, (c) 2-photon subtracted thermal state, (d) single-photon and (e) heralded 9-photon state that emulates emission from a cluster of single-photon emitters. Inset: Wigner function evaluated from the measured statistics. Note that data agree with theory even beyond the number of channels of the PNRD (ten).}
\label{fig_scheme}
\end{figure*}

Contemporary PNRD technologies all rely on multiplexing with the exception of transition-edge detectors \cite{Nam2016,Silberhorn2016} that require temperatures below 100\,mK, offer rates of 10-100 kHz and suffer from range-versus-crosstalk compromise.
Photon-number resolution using a single superconducting nanowire single-photon detector (SNSPD) is also possible, but suffers from significant crosstalk \cite{Cahall2017Dec}.

The multiplexing approach is based on dividing an input optical signal into multiple on-off detectors \cite{Jex1996,Vogel2012}.
Many schemes of temporal and spatial multiplexing have been reported using bulk on-off detectors \cite{Banaszek2003,Walmsley2003,Franson2003,Hradil2003,Silberhorn2018}, integrated on-off pixels \cite{Fiore2008,Andreoni2010,Krivitsky2011,Fiore2016,Hage2017,Berggren2018}, or even a few photon-number resolving detectors \cite{Nam2013b,Silberhorn2016}.
Though being economical in respect to the number of on-off detectors employed, the temporal multiplexed scheme trades off a decrease of the detector speed for an increase in a number of the detection channels. Decreasing the losses and the balancing of temporal multiplexers require a great deal of optimization \cite{Jezek2008} or even active signal switching \cite{Silberhorn2018}.
On the other hand, multiple-pixel PNRDs typically suffer from strong crosstalk effects, which demands an extensive characterization of the detector \cite{Walmsley2008} and
advanced numerical post-processing to correct for the imperfections \cite{Krivitsky2011,Hage2017}. Also, the multi-pixel detectors offer very limited reconfigurability and complicate channel balancing. Recent on-chip integration of independent on-off cryogenic detectors represents a promising direction \cite{Pernice2015,Fiore2016,Berggren2018}, which has yet to be tested for various classical and, particularly, non-classical sources.

The reported photon-number-resolving detector is based on spatial multiplexing of the input photonic signal by a reconfigurable optical network as depicted in Fig.~\ref{fig_scheme}(a). The multiport network consists of cascaded tunable beam splitters composed of a half-wave plate and a polarizing beam splitter, which allow for accurate balancing of the output ports or, if needed, changing their number so there is no need to physically add or remove detectors. The whole network works as a 1-to-$M$ splitter balanced with the absolute error below $0.3\%$.
To measure the multiplexed signal we use single-photon avalanche photodiodes (SPADs) with efficiency close to 70\%, 250~ps timing jitter, and 25~ns recovery time.
The electronic outputs of the SPADs are summed by a custom coincidence logic while keeping the individual channels synchronized. Alternatively, the output can be visualized using an oscilloscope, see Fig.~\ref{fig_scheme}(a). Each of the resulting $M+1$ distinct voltage levels corresponds to the particular number of $m$-fold coincidences. Repeated measurements give rise to click statistics. Full technical details are given in Supplemental Material, including a discussion of processing electronic signals from single-photon detectors.

It is important to stress here that the PNRD operates in real time and yields a result for every single input pulse with a latency (input-output delay) lower than 30 ns including the response of the SPADs, which allows its application also as a communication receiver, quantum discrimination device, or for a feedback operation.
The use of independent detectors and well-balanced coincidence circuitry completely removes any crosstalk between the histogram channels, see Fig.~\ref{fig_scheme}(a). The effects of dark counts and afterpulses are virtually eliminated by operating the detector in pulse regime with the repetition rate below 5 MHz. 
The period between individual measurement runs can be ultimately decreased to be only slightly longer than the recovery time of the constituent single-photon detectors, provided that afterpulsing is low enough.
Furthermore, differences in SPAD efficiencies and other optical imperfections or imbalances of the PNRD can be arbitrarily compensated by adjusting the splitting ratios of the optical network. The result is a balanced multiplex with an overall efficiency $\eta$.
This means that all systematic errors are eliminated either by design or by a sufficiently precise adjustment, independently of constituent detectors employed.



For a balanced $M$-channel PNRD with efficiency $\eta$, the probability of $m$ channels clicking upon the arrival of $n$ photons is

\begin{equation}
C_{mn} =
\binom{M}{m} \sum_{j=0}^{m}\left ( -1 \right )^{j} \binom{m}{j} \left [ \left ( 1-\eta  \right )+\frac{\left ( m-j \right )\eta }{M} \right ]^{n}.
\end{equation}
The click statistics $c_m$ is then determined by the photon statistics $p_n$ \cite{Jex1996,Franson2003,Vogel2012},

\begin{equation}\label{click}
c_{m} = \sum_{n}  C_{mn} p_{n}.
\end{equation}

Finding the photon statistics $p_n$, $n=0\ldots \infty$, that satisfies the system of equations (\ref{click}) for a measured click statistics $c_m$, $m=0\ldots M$, represents the core problem of photon statistics retrieval. This generally ill-posed problem suffers from underdetermination and sampling error. Fortunately, we have additional constraints facilitating the retrieval, i.e. the photon-number probabilities are non-negative, normalized, and typically non-negligible only within a finite range.

Here we present a novel approach, termed expectation-maximization-entropy (EME) method, based on an expectation-maximization iterative algorithm weakly regularized by a maximum-entropy principle. The initial zeroth iteration is uniform; $p_n^{(0)} = 1/(n_{\text{max}}+1)$ for sufficiently large $n_{\text{max}} \gg \langle n \rangle$. Each subsequent iteration is
\begin{equation}\label{EME1}
p_n^{(k+1)} = \Pi_{n}^{(k)} p_n^{(k)} - \lambda \left( \ln p_{n}^{(k)} - S^{(k)} \right) p_{n}^{(k)},
\end{equation}
\begin{equation}\label{EMEkernel}
\Pi_{n}^{(k)} = \sum_{m=0}^{M} \frac{c_m}{\left( \sum_{j}  C_{mj} p_j^{(k)} \right)} C_{mn}\,, \quad
S^{(k)} = \sum_{n=0}^{n_{\text{max}}} p_n^{(k)} \ln p_n^{(k)}.
\end{equation}
Here the superscript $(k)$ denotes $k$-th iteration. Each iteration is evaluated for $n = 1,\ldots,n_{\textrm{max}}$.
The $\Pi_{n}^{(k)}$ is a function of the measured click statistics $c_m$ and the efficiency $\eta$ determined by a separate measurement. $S^{(k)}$ is a negative von Neumann entropy.
The parameter $\lambda$ scales the entropy regularization relative to the likelihood maximization; we use a fixed value of $10^{-3}$ for all the photon statistics.
The process is stopped when two subsequent iterations are practically identical. The retrieved statistics does not change for different initial iterations.
The derivation of the algorithm is given in the Supplemental Material.

To show the accuracy and the robustness of the novel EME method, a numerical analysis was performed for 25 various photon statistics with different mean photon numbers. We compared the EME method with other frequently used algorithms---direct inversion and the expectation-maximization (EM) method based on likelihood maximization. EME was found to be a unique estimator that guarantees non-negativity and the absence of numerical artifacts in the retrieved photon statistics. Total variation distance $\Delta=\sum_n|p_n-p_n^{\textrm{\scriptsize true}}|/2$ between the retrieved distribution and the true one is in the order of $\sim 10^{-3}$, one order of magnitude smaller than in the case of direct inversion and maximum-likelihood approaches. Numerical simulations yield average fidelity values $\overline F=0.9996$ using the EME algorithm and $\overline F=0.997$ using the maximum-likelihood approach.
The fidelity, defined as $F=\textrm{Tr}[(p_n\cdot p_n^{\textrm{\scriptsize true}})^{1/2}]^2$, cannot be evaluated for direct inversion due to negative values of estimated photon statistics.

\begin{figure}[t]
\centering
\includegraphics[width=1.0\columnwidth]{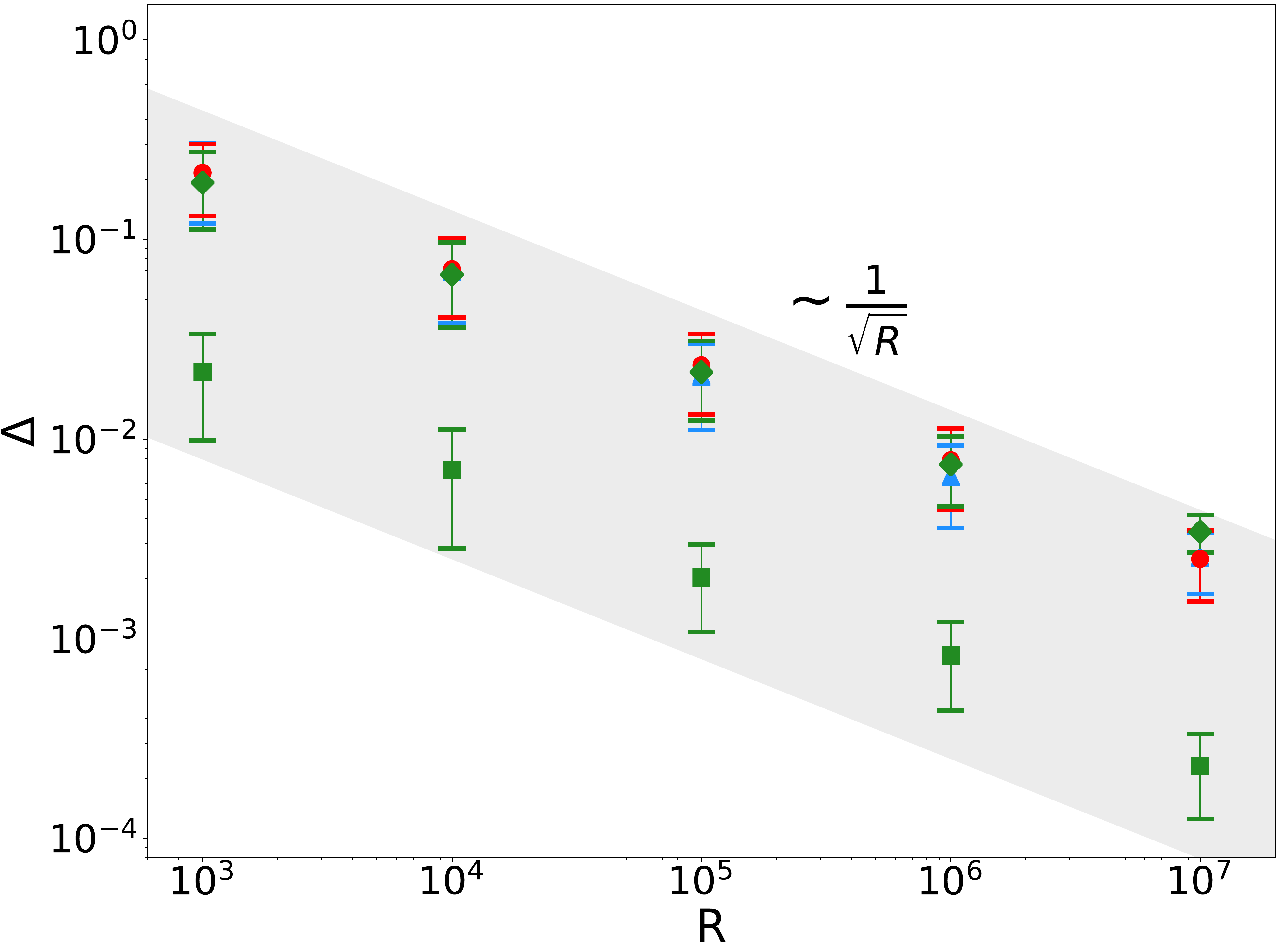}
\caption{A numerical analysis of EME total variation distance $\Delta$. With more measurement runs $R$, the statistical error in the data is lower and the EME result approaches the true photon statistics despite the limited number of channels $M=10$. Here shown for various photon-number distributions and a single value of  $\lambda = 10^{-3}$. Shown are: coherent state with $\langle n\rangle = 10$ (blue triangle up), thermal state with $\langle n\rangle = 5$ (red circle), $N_p$-photon cluster with $N_p=1$ (green square), and $N_p=9$ (green rhombus). The gray area illustrates the observed scaling ($0.25/\sqrt{R}$ to $14/\sqrt{R}$.)}
\label{fig_dataset}
\end{figure}

In Fig. \ref{fig_dataset}, we show by numerical simulation that the results of the EME algorithm approach the respective theoretical expectations as more data is acquired. This means that despite a limited number of channels $M=10$, the chief source of error is the statistical/sampling error.
We also verified that $\Delta$ stays the same if both the mean number of photons and the number of channels are doubled.
Therefore, EME scales well to high photon numbers considering limited experimental resources.
The precision of the photon statistics retrieval can be further increased by optimizing over multiple parameters, such as $M$, $\lambda$, $n_{\textrm{max}}$, or iteration cut-off. Eventually, $\Delta$ becomes limited by machine precision and computation time. The analysis of the complex interaction of these parameters will be the subject of further research.
We also found that the EME convergence is 10--1000$\times$ faster than the plain EM approach (see the Supplemental Material), while yielding significantly better results.

\begin{table*}[ht!]
	\centering
	\begin{tabular}{|c|c|c|c|c|c|c|c|c|c|c|}
\hline
\multirow{2}[4]{*}{} & \multicolumn{2}{c|}{coherent} & \multicolumn{2}{c|}{thermal} & \multicolumn{2}{|c|}{\begin{tabular}[c]{@{}c@{}}2-photon \\ subtracted thermal\end{tabular}} & \multicolumn{2}{c|}{single photon} & \multicolumn{2}{c|}{9-photon cluster} \\
\cline{2-11}      & EM    & EME   & EM    & EME   & EM    & EME   & EM    & EME   & EM    & EME \\
\hline
$F$   & 0.6(1) & 0.9984(9) & 0.69(2) & 0.9978(5) & 0.90(1) & 0.9990(4) & 0.99394(2) & 0.99912(1) & 0.5467(2) & 0.99930(2) \\
\hline
$\Delta$ & 0.50(9) & 0.002(9) & 0.35(1) & 0.033(3) & 0.21(1) & 0.019(6) & 0.07424(1) & 0.00088(1) & 0.1752(5) & 0.00407(1)\\
\hline
\end{tabular}%
\caption{The comparison of EM and EME results for the measured data. Coherent state $\langle n\rangle = 4.95(2)$, thermal state $\langle n\rangle = 4.93(4)$, 2-photon subtracted thermal state $\langle n\rangle = 5.77(2)$, single photon state $N_{p}=1$, and 9-photon cluster $N_{p}=9$. Both fidelity $F$ and total variation distance $\Delta$ are shown. Standard deviations are evaluated by repeating the measurement and data processing ten times. The large distances observed for EM stem from overfitting the ill-posed problem. This is discussed in the Supplemental Material.}
\label{table_review}
\end{table*}

In our experimental demonstration, we used a balanced 10-channel configuration of the detector. We analyzed coherent states, thermal states, multi-mode thermal states, single-photon and multiple-photon subtracted thermal states, and non-classical multiphoton states. Furthermore, we have varied the mean number of photons, the number of modes, and the number of subtracted or superimposed photons. For each retrieved photon statistics we computed $\langle n\rangle$, $g^{(2)}$, and other quantities presented in detail in the Supplemental Material.

The measurements were performed using 1-ns-long optical pulses with the repetition rate of 2 MHz. We prepared the initial coherent signal by using a gain-switched laser diode at 810 nm. The resulting coherent pulses measured by the PNRD show almost perfect Poissonian statistics with $g^{(2)} = 1$ up to $\langle n\rangle = 20$ with average fidelity $\overline{F}=0.996$ and total variation distance $\overline{\Delta}=24\times{10^{-3}}$.
The thermal state is generated by temporal intensity modulation of the initial coherent light by a rotating ground glass. The scattered light is collected using a single-mode optical fiber.
We measured almost ideal Bose-Einstein photon statistics depicted in Fig.~\ref{fig_scheme}(b) with $g^{(2)} = 2$ up to $\langle n\rangle = 5$, $\langle\Delta n^2\rangle = 30$ with $\overline{F}=0.997$ and $\overline{\Delta}=24\times10^{-3}$.
We varied the number of the collected thermal modes, which yielded a signal governed by Mandel-Rice statistics, going from Bose-Einstein to Poisson distribution as the number of modes increased.
Multiple-photon subtraction from the thermal state was implemented using a beam splitter with a $5\%$ reflectance. When a (multi)coincidence was detected by a multichannel single-photon detector in the reflected port, the heralded optical signal in the transmitted port was analyzed by the reported PNRD. A typical result of 2-photon subtraction is shown in Fig.~\ref{fig_scheme}(c). Increasing the number of subtracted photons results in a transition from super-Poissonian chaotic light to Poissonian signal \cite{Olivares2010,Migdall2013}.
Furthermore, we generated multi-photon states by mixing incoherently several single-photon states from spontaneous parametric down-conversion using time multiplexing.
$N_p$ successive time windows, where a single photon was heralded, were merged into a single temporal detection mode. This source emulates the collective emission from identical independent single emitters \cite{molecules2,Hell2017,Straka2018}.
The resulting photon statistics measured for these highly nonclassical multi-photon states corresponds extremely well to the ideal attenuated $N_p$-photon states, see Fig.~\ref{fig_scheme}(d,e) for $N_p=1$ and $9$ with $\overline{F}=0.999$ and $\overline{\Delta}=3\times10^{-3}$.
Also the $g^{(2)}$ parameter computed from the measured photon statistics perfectly agrees with the theoretical model $1 - 1/N_p$, see Fig.~\ref{fig_g2_vs_mean}.

We utilize fidelity and total variation distance to compare the measured distribution with the ideal one. The worst and the best fidelities $F=0.985$ and $0.9999$ are reached across all the tested sources with average fidelity being $\overline F=0.998$.
The average distance is $\overline\Delta = 17\times10^{-3}$ for all the tested sources. For detailed data and comparison to plain EM, see Table~\ref{table_review} and the Supplemental Material.
The errors of EME are caused by slight imbalances of splitting ratios in the PNRD, variations in PNRD efficiency $\eta$, and imperfections of the tested sources, which renders the actual accuracy of the PNRD even higher. Particularly, accurate preparation and characterization of thermal and super-chaotic states are highly nontrivial tasks subject to ongoing research \cite{Bondani2015,Saleh2016,Slodicka2018,Chekhova2018rogue}.


To conclude, we have reported a fully reconfigurable near-ideal photon-number-resolving detection scheme with custom electronic processing and a novel EME photon statistics retrieval method.
The PNRD design is free of systematic errors, which are either negligible or can be arbitrarily decreased by the user. We have demonstrated exceptional accuracy of detected photon statistics that goes beyond the conventional limit of the number of PNRD channels. We measured dozens of various photonic sources ranging from highly non-classical quantum states of light to chaotic optical signals. The results were obtained from raw data with no other processing than EME, and without any demanding detector characterization. Despite uncorrected systematic errors and significant variability of the input signal, our approach shows superior fidelity across the board with typical values exceeding $99.8\%$ for mean photon numbers up to 20 and the $g^{(2)}$ parameter reaching down to a fraction of a percent. Though having been demonstrated with common single-photon avalanche diodes, the reported measurement workflow is independent of the detection technology and can accommodate any on-off detectors. Furthermore, the multi-channel scheme allows for straightforward on-chip integration. Therefore, further improvements in speed, efficiency and compactness can be expected using superconducting single-photon detectors \cite{Fiore2008,Fiore2016,Berggren2018} coupled with waveguide technology \cite{Fiore2013,Szameit2016,Pernice2018}.

\nocite{Note2}

\section*{Acknowledgments}
We acknowledge the support from the Czech Science Foundation under the project 17-26143S.
This work has received national funding from the MEYS and the funding from European Union's Horizon 2020 (2014-2020) research and innovation framework programme under grant agreement No 731473 (project 8C18002). Project HYPER-U-P-S has received funding from the QuantERA ERA-NET Cofund in Quantum Technologies implemented within the European Union's Horizon 2020 Programme. We also acknowledge the support from Palack\'y University (projects IGA-PrF-2018-010 and IGA-PrF-2019-010).

\clearpage
\onecolumngrid
\appendix

\centerline{\large\bf{Supplementary Material}}

\section{Characterization of quantum states of light with focus on their photon statistics}

The photon statistics $p_n$ retrieved from a click statistics recorded by the photon-number-resolving detector (PNRD) can be analyzed and compared with a theoretical model but very often only a limited number of characteristics are evaluated and utilized to witness a particular feature of the source under the test.
The features crucial for fundamental research as well as many photonic applications are non-classicality and a deviation from Poisson statistics.
The normalized second-order intensity correlation $g^{(2)}=\langle n(n-1)\rangle/\langle n\rangle^2$ is routinely applied to quantify the non-classicality of light \cite{Mandel1977,Saleh2016}.
The departure of the photon statistics from Poisson distribution is frequently characterized by the Mandel parameter $Q=[\langle(\Delta n)^{2}\rangle - \langle n\rangle]/\langle n\rangle$ \cite{Mandel1979,OlayaCastro2014}. Evaluation of these parameters requires the full knowledge of the photon statistics, though an approximate value of $g^{(2)}$ can be measured directly for $\langle n\rangle\ll1$ \cite{Grangier1986,Bjork2018}. Both characteristics depend on the first two moments of photon statistics, particularly the mean photon number $\langle n\rangle$ and the second moment $\langle n^2\rangle$, but behave differently in the presence of losses. The normalized second-order correlation is completely loss independent while the Mandel $Q$ parameter changes with the applied losses. For example, the multi-photon states emitted by clusters of single-photon emitters would ideally display $Q=-1$, however, the real sources with limited collection efficiency show the value closer to zero. In our case of the emission emulated by merging of several heralded photons from parametric down-conversion process, the measured value $Q=-0.548(2)$ is given by the heralding efficiency of the source.

Quantum non-Gaussian (QNG) states represent another class of strongly non-classical states, which cannot be expressed as a mixture of arbitrary Gaussian states. We were able to prove QNG of the measured multi-photon states up to nine photons \cite{Straka2018} using reconfigurability of our PNRD detector and applying recently devised click-based QNG criteria \cite{Lachman2016}.
Having the full photon statistics retrieved, we can also apply QNG tests based on state properties in phase space \cite{Paris2013,Park2015,Park2017,Schleich2018}. Unfortunately, they rely on a single-mode assumption, which does not hold for our multi-photon states emulating a superimposed emission from clusters of single-photon emitters. Anyway, we have successfully proved QNG property for the emission from the cluster of 1, 2, and 3 emitters using the criterion by Genoni et al. \cite{Paris2013}.
QNG criteria feature strong sensitivity to losses compared to non-classicality measures \cite{Straka2014}.

Even higher loss sensitivity can be observed using the mean value of parity operator $\langle {\cal P}\rangle = \sum_n (-1)^n p_n$ and the Wigner function $W(x,p) = \sum_n p_n W_n(x,p)$, where $W_n(x,p) = \int_{-\infty}^{\infty} \!{\textrm{d}y} \, \psi_{n}(x-y) \psi_{n}^{*}(x+y) \textrm{e}^{2iyp}/\pi$ and $\psi_n(x)$ is a wave function of the $n$-th Fock state. The negativity of $\langle {\cal P}\rangle$ follows the negativity of Wigner function at the origin of phase space as $W(0,0) = \langle {\cal P}\rangle/\pi$, which we have detected for the single-photon state (emission from a single emitter) and the cluster of three single emitters.

In the main text, we focus mainly on the discrepancy between the measured and the corresponding ideal photon statistics for dozens of measured photonic sources, characterized by the fidelity ${\cal F}=\textrm{Tr}[\sqrt{p_n\cdot p_n^{\textrm{\scriptsize ideal}}}]^2$ and the total variational distance $\Delta=\sum_n|p_n-p_n^{\textrm{\scriptsize ideal}}|/2$, both ranging from 0 to 1.
Higher fidelity does not necessarily correspond to a stronger non-classical feature so other characteristics should also be evaluated \cite{Paris2016}.
We demonstrate the accuracy and wide applicability of the PNRD by plotting $g^{(2)}$ parameter computed from the measured photon statistics and from the corresponding ideal statistics for 25 various optical signals, see Fig.~1 of the main text.
Here in the Fig.~\ref{fig_Mandel_vs_mean} of Supplemental material we demonstrate a similar characterization utilizing the Mandel $Q$ parameter. We can see the exceptionally accurate experimental characterization of the deviation from Poissonian statistics with the Mandel $Q$ parameter ranging from -0.55 to 5 and the mean photon number exceeding 20.
Furthermore, we show all the mentioned moments, non-classicality parameters, and Wigner-function-negativity characteristics for five selected photonic signals, namely the four signals, the photon statistics of which is plotted in Fig.~2 of the main text, and the initial coherent state, see Table~\ref{table_review}.

\begin{figure}[!h!]
	\centering
	\includegraphics[width=0.48 \columnwidth]{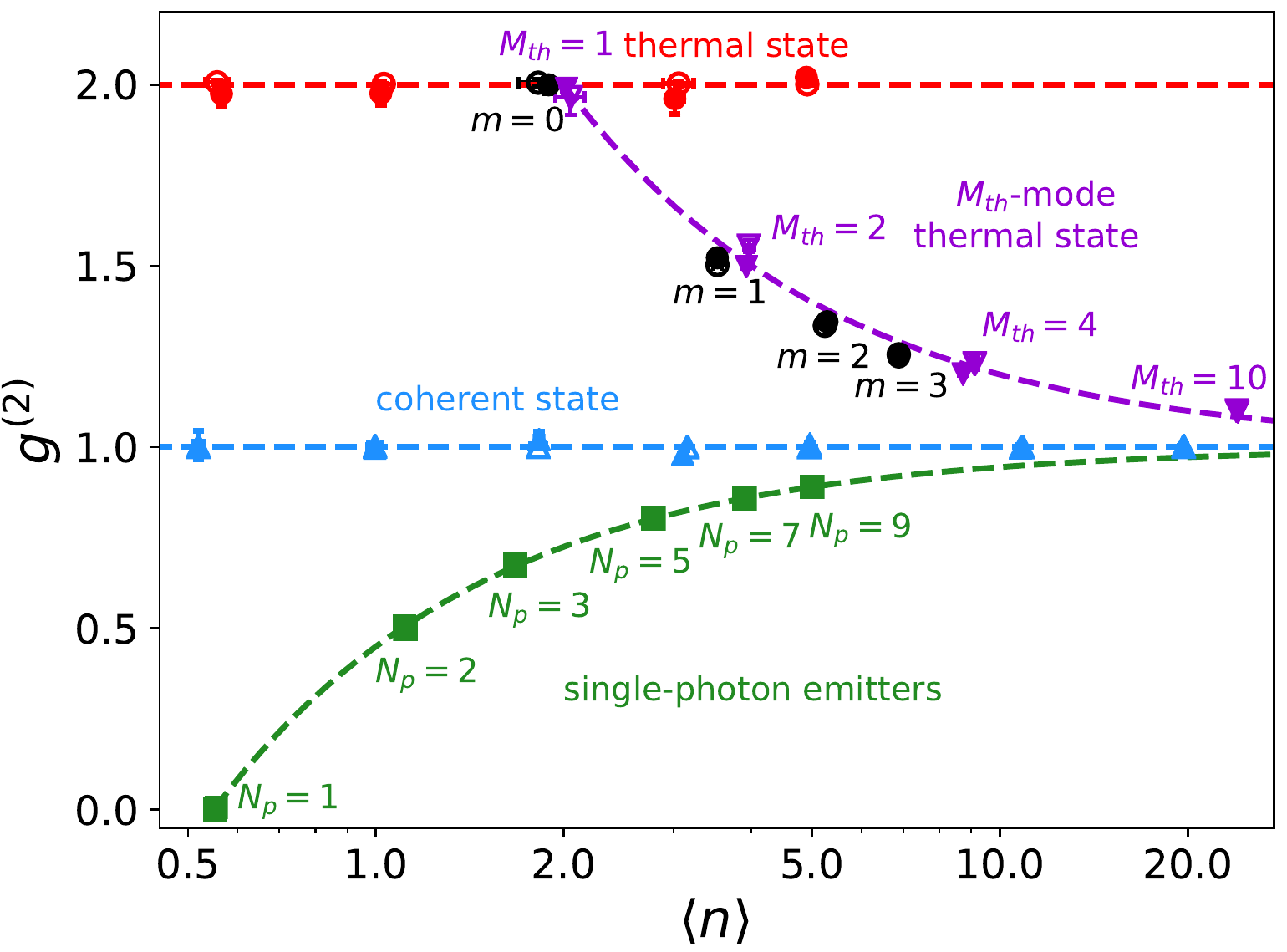}\hspace*{5mm}\includegraphics[width=0.48 \columnwidth]{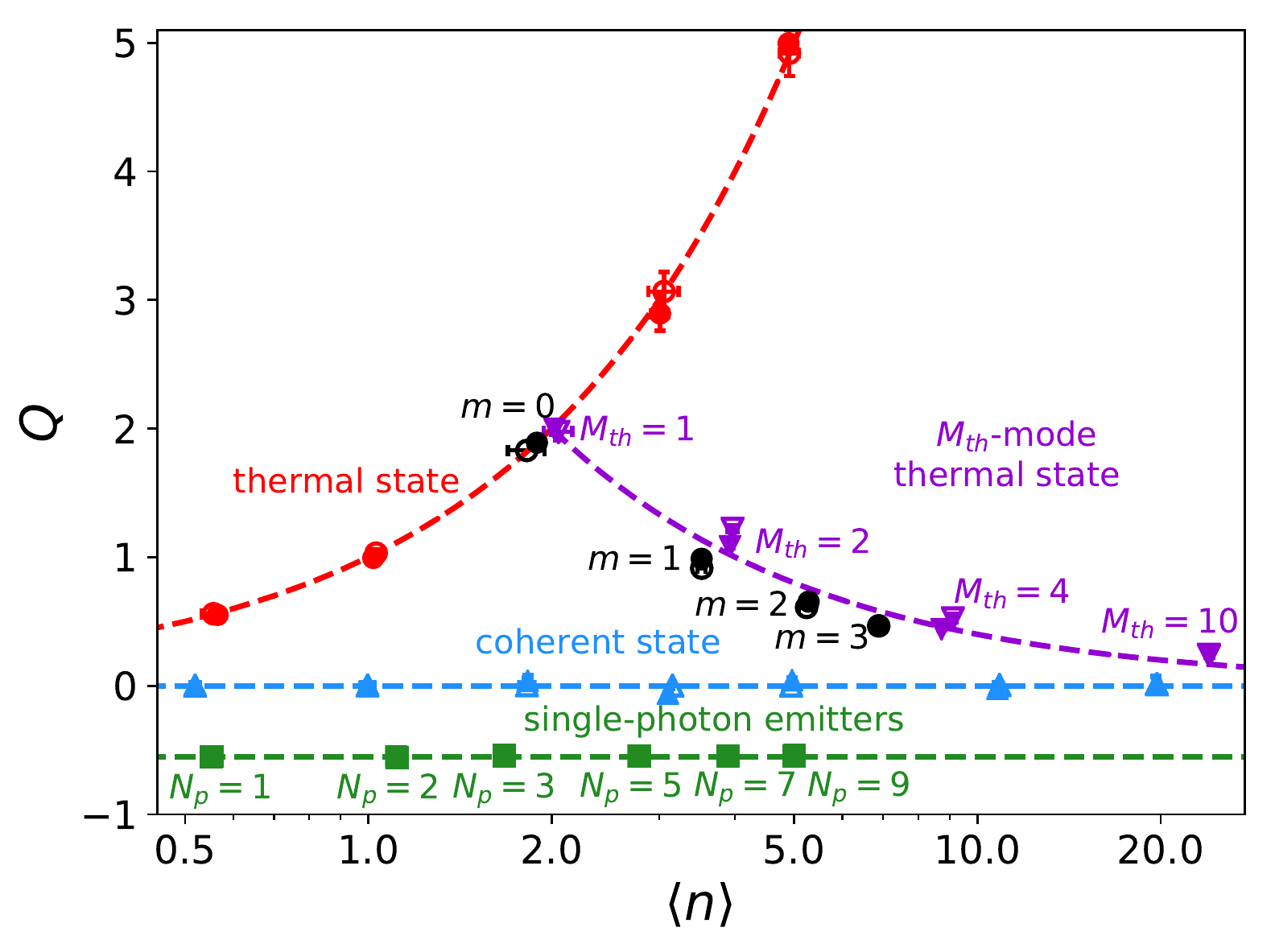}
	\caption{{\em The left sub-plot} is reproduced from the main text: the normalized second-order intensity correlation $g^{(2)}$ evaluated from the measured photon statistics (solid marker) and the corresponding ideal statistics (empty marker) of various optical signals with mean photon number $\langle n\rangle$. Shown are: coherent states with $g^{(2)}=1$ (blue triangle up), thermal states (also termed chaotic light) with $g^{(2)}=2$ (red circle), $M_{th}$-mode thermal states with $M_{th}=1,2,4,10$ (violet triangle down), and $m$-photon-subtracted thermal states for $m=0,1,2,3$ (black circle). The cases of $M_{th}=1$ and $m=0$ coincides with thermal state. Furthermore, the emission from a cluster of $N_p$ single-photon emitters is shown for $N_p=1\ldots9$ with $g^{(2)}=1-1/N_p$.
		{\em The right sub-plot} shows the Mandel $Q$ parameter evaluated from the measured photon statistics of the same optical signals with the same marker/color coding. Coherent states are compatible with Poisson statistics $Q=0$, thermal states show super-Poissonian statistics with $Q=\langle n\rangle$, multi-mode thermal states and multi-photon-subtracted thermal states converges to Poisson distribution with increased number of modes/subtractions. Highly nonclassical $N_p$-photon states reach $Q=-0.55$ given by the limited efficiency of the single-photon source employed for their generation.
		The error bars are typically smaller than the symbol size.}
	\label{fig_Mandel_vs_mean}
\end{figure}

\begin{table}[!h!]
	\begin{center}
		\begin{tabular}{|c|c|c|c|c|c|c|c|c|c|c|}
			\hline
			& \multicolumn{2}{c|}{coherent} & \multicolumn{2}{c|}{thermal} & \multicolumn{2}{|c|}{\begin{tabular}[c]{@{}c@{}}2-photon \\ subtracted thermal\end{tabular}} & \multicolumn{2}{c|}{single photon} & \multicolumn{2}{c|}{9-photon cluster}  \\
			\hline
			& data  & model & data  & model & data  & model & data  & model & data  & model \\
			\hline
			$\left \langle n \right \rangle$ & 4.95(2) & 4.95  & 4.93(4) & 4.93  & 5.77(2) & 5.77  & 0.554675(2) & 0.55  & 5.00786(2) & 4.95 \\
			\hline
			$\langle(\Delta n)^{2}\rangle$ & 4.99(5) & 4.95  & 29.4(5) & 29.24 & 17.0(2) & 16.87 & 0.2487836(3) & 0.2475 & 2.27821(7) & 2.2275 \\
			\hline
			$g^{(2)}$ & 1.002(3) & 1.0   & 2.01(1) & 2.00  & 1.336(3) & 1.333 & 0.0057627(9) & 0     & 0.891156(3) & 0.8889 \\
			\hline
			$Q$   & 0.01(1) & 0.0   & 4.97(7) & 4.93  & 1.94(2) & 1.92  & -0.551478(2) & -0.55 & -0.54507(1) & -0.55 \\
			\hline
			$\left \langle P \right \rangle$ & -0.003(8) & $5\times10^{-22}$ & 0.089(7) & 0.092 & 0.011(5) & 0.00081 & -0.105814(4) & -0.1  & -     & - \\
			\hline
			$W(0,0) $ & -0.001(3) & $2\times10^{-22}$ & 0.028(2) & 0.029 & 0.004(2) & 0.00026 & -0.033682(1) & -0.03184 & -     & - \\
			\hline
			
		\end{tabular}%
		
	\end{center}
	\caption{Characteristics of measured photon statistics for selected states: coherent state, thermal state, 2-photon subtracted thermal state, 1-photon state, and 9-photon mixture. Standard deviations are evaluated by repeating ten times the whole process of the PNRD measurement, photon statistics retrieval, and characteristics evaluation. The amount of data acquired for single-emitter clusters were several orders of magnitude larger that for other states, which yields the corresponding error bars much smaller.}
	\label{table_review}
\end{table}

\section{Experimental setup of the detector, data acquisition and processing}

The reported photon-number-resolving detector (PNRD) is based on spatial multiplexing in a reconfigurable optical network as depicted in Fig.~2(a) of the main text. The multiport optical network consists of tunable beam splitters composed of a half-wave plate and a polarizing beam splitter for accurate adjustments of the splitting ratio. The half-wave plates can be adjusted to split the light equally to any number of output channels, ten in our case, so there is no need to physically add or remove detectors.
Each of the 10 channels is coupled to a multimode fiber and brought to a single-photon avalanche photodiode (SPAD, Excelitas) with system efficiency ranging from 60 to 70\% at 810 nm, 200-300~ps timing jitter, and 20-30~ns dead time.
Different efficiencies of the SPADs are taken care of during balancing. The splitting ratios are set so that the detection rate in each channel is the same. As a result, the overall transmittance of each channel is the same (the product of the optical transmittance of the particular port and the efficiency of the SPAD sitting in that port). The PNRD then becomes a balanced detection multiplex with a global efficiency $\eta$ that is a combination of all constituent losses.

The electronic outputs of the SPADs are processed by a custom coincidence logic. We have tested two implementations of the coincidence circuit, analog and digital ones, with the virtually same results. The analog solution employs commercial electronic modules, namely 300 MHz discriminators (Phillips Scientific NIM MODEL 708), delay lines (Phillips Scientific NIM MODEL 792), and an array of 250 MHz linear fan-in/out units (Phillips Scientific NIM MODEL 740). The propagation delay is approximately 10 ns (excluding coaxial patch cords) and the coincidence window should be larger than 20 ns because of few-nanosecond rise and fall times and time jitter. The bandwidth can be further increased utilizing a passive RF summation circuitry (Mini Circuits ZC16PD-252-S+) instead of the active fan-in units. We verified this option and reached the propagation delay below 5 ns, coincidence window 10 ns, and sub-ns jitter given mainly by the discriminator. The number of channels of the PNRD can be increased to several dozens while keeping the same analog electronic signal processing technique. Potential disbalance in the summation circuitry can be corrected by careful adjusting the amplitude of the individual electronic pulses produced by the discriminator. After the analog summation, the output signal is digitized with 20 GSa/s by a 1.5~GHz oscilloscope operating in a memory-segmentation regime (Teledyne LeCroy). Each of the thresholded voltage levels (eleven in our case) corresponds to the particular number of multi-coincidences.

Alternatively, a fully digital coincidence device can be employed using either TTL logic in field programmable gate arrays (FPGAs) or discrete ECL logic gates. We developed a custom 16-channel emitter-coupled logic (ECL) circuitry consisting of fast comparators, delay lines and basic gates, and demonstrated the coincidence window within a range of 600 -- 12,000 ps with 10 ps timing resolution and 20 ps overall jitter. The propagation delay of the whole coincidence unit is below 5 ns. The single-run output of the unit is stored in 16 flip-flops gates and can be either used directly for fast electro-optic feedforward or read by a microcontroller and sorted in a coincidence histogram for repeated measurements. The developed device performs a real-time classification of all possible detection events in $2^{16}$-element histogram with the rate up to 4 million events per second, with possible increase by order of magnitude using FPGA instead of the microcontroller. The full histogram contains more information than required for click statistics and is further reduced to just 17 elements -- no detection, singles, two-photon events,... 16-photon events.

\begin{figure*}[!ht!]
	\centering
	\includegraphics[width=0.95\textwidth]{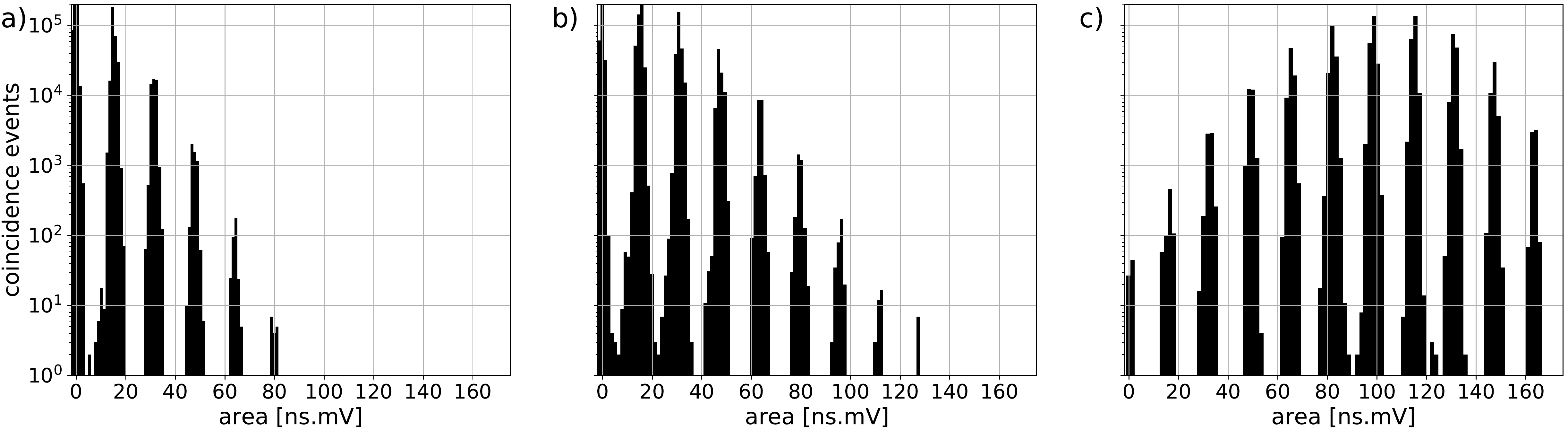}
	\caption{Pulse-height spectrum of the analog output of the reported PNRD for coherent states of various mean photon number $\langle n\rangle=$ 1 (a), 5 (b), and 20 (c). The spectra are plotted in log-scale to emphasize the perfect energy resolution and the absence of any crosstalk effects or background noise.}
	\label{fig_spectrum}
\end{figure*}

The use of independent detectors and well-balanced coincidence circuitry removes completely any crosstalk between the histogram channels yielding the perfect energy quanta resolution up to number of the channels used, see Fig.~\ref{fig_spectrum}. Furthermore, the effects of dark counts and afterpulses are virtually eliminated by operating the detector in pulse regime with the repetition rate below approx{.} 5 MHz \cite{Polyakov2017,Ursin2018}. The period between individual measurement runs can be ultimately decreased to be only slightly longer than the recovery time of the constituent single-photon detectors, as far as afterpulses are negligible or fast decaying like in the case of superconducting nanowire single-photon detectors \cite{Sasaki2011,Hadfield2013,Pernice2018}. Consequently, the presented PNRD measurements are free of systematic errors such as the channel crosstalk and temporal correlations.

\section{Photonic sources characterized using the PNRD detector}

An initial coherent nanosecond pulsed light is generated by gain-switched semiconductor laser diode (810 nm) driven by sub-nanosecond electronic pulse generator with a repetition rate of 2 MHz.
We have reached $<1\%$ pulse-to-pulse stability and similar long-term stability of the mean power by employing custom low-noise low-jitter pulse generator, laser diode selection, its thermal stabilization, and optimization of driving pulse duration and shape. The resulting coherent pulses measured by the PNRD detector display virtually perfect Poissonian statistics.

Thermal states are generated by temporal intensity modulation of the initial coherent light by rotating ground glass (RGG) with a random spatial distribution of speckles \cite{Spiller1964}.
A single-mode optical fiber is used to collect the scattered light and select a single spatial mode. We have measured the corresponding photon statistics to be nearly ideal Bose-Einstein distribution.
Let us note that it is quite challenging to reach an ideal thermal state, see \cite{Slodicka2018} and reference therein. In the initial stage of our experiment, we observed a discrepancy between measured photon statistics of RGG modulated light and the ideal Bose-Einstein statistics. It appeared later that the error was caused by a small inhomogeneity of the RGG disk. More consistent results can be attained using a direct programmable modulation of light intensity, which allows for preparation of near-ideal thermal state and also an arbitrary photon statistics \cite{Straka2018generator}.

We have altered the effective number $M_{th}$ of thermal modes by changing the size of the speckles collected via single-mode fiber. This was achieved by changing the diameter of the laser spot on the RGG disk and the distance between the disk and the fiber tip. The resulting Mandel-Rice statistics changes from Bose-Einstein to Poisson distribution with increasing number of modes.
The multi-mode thermal state shows the largest discrepancy between the measured photon statistics and the corresponding ideal one, which is caused by its relatively complicated preparation. The intensity of the initial coherent state, its focusing on the RGG, and the fiber coupling are changed to simultaneously reach the required mean photon number and the variance compatible with the Mandel-Rice statistics, basically verifying $g^{(2)}=\frac{2 M_{th}+1}{M_{th}}-1$, where $M_{th}$ is number of modes. Also, the RGG used has to be checked first for its roughness homogeneity by producing the ideal chaotic light with the Bose-Einstein photon statistics. If the RGG allows generating a super-chaotic statistics for any combination of the mentioned hardware parameters, it cannot be straightforwardly used to produce a proper transition from Bose-Einstein to Poisson statistics via the multi-mode Mandel-Rice statistics. Indeed, several modes of super-chaotic light can yield the total statistics different from the Bose-Einstein distribution but displaying $g^{(2)}=2$.

Multi-photon subtraction from the thermal state is implemented by splitting the thermal state at a beam splitter with a reflectance $R = 5\%$. When a (multi)coincidence is detected by a reconfigurable multi-channel detector (multiple SPADs) in the reflected port, the heralded optical signal in the transmitted port is analyzed by the reported PNRD. With increasing number $m$ of subtractions, the mean photon number of the conditioned output state is $m$-times larger than of the original thermal state and the resulting photon statistics converges to a Poisson distribution. The transition follows a similar path in $\langle n\rangle$--$g^{(2)}$ and $\langle n\rangle$--$Q$ diagrams as multi-mode thermal light, however, the multi-photon subtracted thermal states are single-mode states with many applications in quantum metrology and quantum thermodynamics \cite{Olivares2010,Migdall2013,Walmsley2016,Hlousek2017,Kulik2017,Barnett2018,Lvovsky2018,Kulik2018}.

Furthermore, we generate multi-photon states by mixing $N_p$ single-photon states incoherently, using the process of continuous-wave spontaneous parametric down-conversion in a PPKTP crystal and time multiplexing.
We took $N_p$ successive time windows, when a single photon was heralded, and joined them into a single temporal detection mode. The resulting photon statistics measured for these highly-nonclassical multi-photon states corresponds extremely well with the ideal attenuated $N_p$-photon states up to $N_p=9$.
The measured mean photon number is lower than the number of superimposed photons due to non-unity efficiency of the source, $\eta_{\textrm{source}}=0.55$, which is contributed the efficiency of the heralding detector (65\%), single-mode-fiber collection efficiency (90\%), and spectral filter transmission and other inefficiencies (94\%).
One might conclude from the $\langle n\rangle$--$g^{(2)}$ diagram shown in Fig.~\ref{fig_Mandel_vs_mean} that non-classicality of the multi-photon state is reduced with the increasing number of photons superimposed. Photon statistics of these states are very close to a binomial distribution for all $N_p$ and so they are strongly non-classical and non-Gaussian, as displayed by the Mandel Q parameter or other advanced criteria \cite{Straka2018,Chekhova2018clusters,Lachman2018}.

We have performed approximately $3\times10^{5}$ measurement runs to build a click statistics for classical photon states. The measurement uncertainty has been evaluated by repeating the full acquisition ten times. For non-classical multi-photon states, we have performed the single acquisition with $10^{11}$ measurement runs and use Monte Carlo simulation for uncertainty evaluation. Monte Carlo method has also been used to quantify the statistical errors of retrieved photon statistics, fidelities, and other parameters of interest.

\section{Theoretical description of the PNRD and photon statistics retrieval}

A perfectly balanced $M$-port PNRD with efficiency $\eta$ and free of dark counts, crosstalk, and other imperfections is described by a conditional matrix,
\begin{equation}\label{Cmatrix}
	C_{mn} =
	\binom{M}{m} \sum_{j=0}^{m}\left ( -1 \right )^{j} \binom{m}{j} \left [ \left ( 1-\eta  \right )+\frac{\left ( m-j \right )\eta }{M} \right ]^{n},
\end{equation}
transforming the photon statistics $p_n$, $n=0\ldots\infty$, to {\em theoretical} click statistics $c_m$, $m=0\ldots M$,
\begin{equation}\label{click}
	c_{m} = \sum_{n=0}^{\infty}  C_{mn} \, p_{n}.
\end{equation}
The element $C_{mn}$ describes the probability of $m$ detected simultaneous clicks conditioned by $n$ incident photons, $C_{mn}=0$ for $m>n$ \cite{Jex1996,Franson2003,Vogel2012}.
The structure of the matrix can be easily understood by representing the actual $M$-port PNRD with non-unity detection efficiency as $(M+1)$-port device with a virtual ``sink'' output with the probability of $(1-\eta)$ and $M$ equiprobable outputs with the overall probability of $\eta$.
The formulation (\ref{click})-(\ref{Cmatrix}) is valid for any input state of light.
The efficiency $\eta$ is inferred from an independent measurements, and is assumed to be constant during the PNRD operation. The simple $C$ matrix (\ref{Cmatrix}) with the same efficiency is used for photon statistics retrieval for all the sources characterized, without the need for a tomographic characterization of the detector \cite{SanchesSoto1999,Fiurasek2001,Walmsley2008,Hradil2010,Hadfield2013tomo,Barbieri2016,Smith2014}.

Finding the photon statistics $p_n$, $n=1\ldots\infty$, that satisfies the system of equations (\ref{click}) for a particular {\em measured} click statistics $d_m$, $m=0\ldots M$, represents a core problem of photon statistics measurement. This inverse problem is ill-posed because 1. it is obviously underdetermined, 2. the theoretical click probabilities $c_m$ are not available in real measurement as we acquire relative frequencies $d_m$ instead, which sample the true probabilities (sampling noise), and 3. for PNRDs that are not free of systematic errors, other imperfection can be present like imbalance, crosstalk, and temporal correlations.
The reported PNRD implementation is almost free of these technical imperfections.
The first two issues, however, remain for any PNRD detector, and the photon statistics retrieval has to take them into account. Fortunately, we have additional information facilitating the retrieval, i.e. the photon number probabilities are non-negative and normalized.
The elements of photon statistics are also typically non negligible only within a finite range of photon numbers. Indeed, the classical states of light possess a quickly decaying tail, and nonclassical states such as single-photon states are actually defined on a finite support. There are many techniques for photon statistics retrieval, direct inversion and maximum-likelihood approach being probably the most frequently employed. In what follows we present the basic ideas of these techniques and present a novel method based on an iteration technique known as expectation-maximization algorithm weakly regularized by maximum-entropy principle.

\textbf{Direct inverse.}
The retrieval technique based on the direct inversion of the system of linear equations (\ref{click}) requires setting a cut-off -- the maximum photon number $n_\textrm{\scriptsize max}$, typically equal to the number of PNRD ports. The truncated problem possesses a single solution
\begin{equation}\label{Inverse}
	\tilde{p}_{n} = \sum_{m=0}^{M} (C^{-1})_{nm} \, d_{m},
\end{equation}
the non-negativity of which is not guaranteed, hence not representing a physically sound photon statistics. Here $C^{-1}$ represents the inverse matrix to the conditional matrix $C$.
The solution (\ref{Inverse}) often reaches negative values and artificial oscillations, see Fig.~\ref{fig_retrieval_methods}. These adverse effects are particularly noticeable in the practical  case of non-unity efficiency $\eta<1$ with the limited number $M$ of output ports and the mean photon number of the incident light comparable or higher than $M$.
The non-negativity constraint can be incorporated using linear programming, for instance, 
which reduces the volume of the $p_n$ domain by the factor of $2^{n_\textrm{\scriptsize max}}$.
Also, the cut-off can be increased $n_\textrm{\scriptsize max}>M$ rendering the problem underdetermined, a pseudoinverse \cite{Rao1971,Ben-Israel2003} of which often diverges or, at least, amplifies a sampling noise. Various regularization methods are used to make these issues less pronounced \cite{Tikhonov1943,Hoerl1962,Tikhonov1995,Gomonay2009}. Despite all the mentioned issues the direct inverse methods are frequently used due to their speed and widespread implementation in many numerical libraries and computing systems.

\textbf{Maximum-likelihood method and expectation-maximization algorithm.}
Another technique to achieve the inversion of the conditional probability matrix is well known maximum-likelihood (ML) principle and the expectation-maximization (EM) algorithm, which provides a robust method for finding a solution (ML estimate) \cite{Dempster1977,Vardi1993}.
The likelihood of measuring the particular data distribution $\{d_m\}$ given the input photon statistics $\{p_n\}$ and measurement device $C$ is is given by the multinomial distribution
\begin{equation}\label{likelihood}
	\prod_{m=0}^M c_m^{d_m}
	= \prod_{m=1}^M \left(\sum_{n=0}^{n_\textrm{\scriptsize max}}C_{mn}p_n\right)^{d_m},
\end{equation}
which is a convex functional defined on a convex set of $\{p_n\}$ distributions. The maximization of the likelihood functional yields a single global maximum in the case of $n_\textrm{\scriptsize max}\leq M$ or a single plateau of maxima in the case of underdetermined problems.
A logarithm of the likelihood is often used instead, which does not change the convexity feature.
Also, the normalization $\sum p_n = 1$ condition is incorporated with the help of a Lagrange multiplier $D$,
\begin{equation}\label{loglikelihood}
	L\left[\{p_n\}\right] = \sum_{m=0}^M d_m \ln c_m - D \left( \sum_{n=0}^{n_\textrm{\scriptsize max}} p_n - 1 \right)
\end{equation}
The zero variation is a necessary condition for an extreme of the likelihood functional,
\begin{equation}\label{variation}
	\delta L = L\left[\{p_n + \delta p_n\}\right] - L\left[\{p_n\}\right]
	= \sum_{m=0}^M \frac{d_m}{c_m} \sum_{n=0}^{n_\textrm{\scriptsize max}} C_{mn} \delta p_n
	- D \sum_{n=0}^{n_\textrm{\scriptsize max}} \delta p_n
	= \sum_{n=0}^{n_\textrm{\scriptsize max}} \left( \sum_{m=0}^M \frac{d_m}{c_m} C_{mn} - D \right) \delta p_n
	= 0
\end{equation}
for each $\{\delta p_n\}$, which is equivalent to
\begin{equation}\label{variation2}
	\sum_{m=0}^M \frac{d_m}{c_m} C_{mn} - D = 0,
\end{equation}
except at the boundary of the domain where $p_n=0$. To include this boundary condition, the extremal equation is formulated as
\begin{equation}\label{extreme}
	\sum_{m=0}^M \frac{d_m}{c_m} C_{mn} p_n = D p_n.
\end{equation}
A summation over $n$  yields
\begin{equation}\label{normalization}
	D = \sum_{m=0}^M \frac{d_m}{c_m} \sum_{n=0}^{n_\textrm{\scriptsize max}} C_{mn} p_n = \sum_{m=0}^M d_m = 1,
\end{equation}
where the constraint $\sum p_n = 1$ and the normalization of the click data have been applied.
The functional (\ref{loglikelihood}) can be maximized over $n_\textrm{max}+1$ variables using downhill simplex method or some other standard numerical method \cite{Banaszek1998ML,Fiore2009}.
To keep the non-negativity constraint the variables $p_n$ can be parametrized as $r_n^2$, the downside of which is even more complicated structure of the log-likelihood function.
This approach is straightforward but numerically demanding as the dimension of the problem increases.
Alternatively, an iterative solution of the extremal equation (\ref{extreme}), which is a form of the EM algorithm, can be carried out as was suggested by Banaszek for the first time \cite{Banaszek1998EM,Hradil2003,Paris2005},
\begin{equation}\label{EM}
	\Pi_{n}^{(k)} p_n^{(k)} = p_n^{(k+1)}, \qquad
	\Pi_{n}^{(k)} = \sum_{m=0}^M \frac{d_m}{\left( \sum_{j} C_{mj} p_j^{(k)} \right)} C_{mn}.
\end{equation}
The iteration process is started with an initial positive statistics, typically chosen as the uniform distribution, $p_n^{(0)} = 1/(n_\textrm{max}+1)$. Then the kernel $\Pi_{n}^{(0)}$ of the map (\ref{EM}) is evaluated for the initial iteration step, and the first iteration $\{p_n^{(1)}\}$ is obtained by the application of the kernel on the initial statistics.
The normalization $p_n^{(1)}/\sum p_n^{(1)}$ has to be performed if the data $\{d_m\}$ are not properly normalized. The iteration process is repeated until the distance between $(k+1)$-th and $k$-th iteration is smaller than some given value,
\begin{equation}\label{stop}
	\sqrt{\sum_n \left( p_n^{(k+1)} - p_n^{(k)} \right)^2} < \epsilon.
\end{equation}
Throughout this work the value $\epsilon=10^{-12}$ is used for all the performed photon statistics retrievals.
When sufficient mathematical conditions are fulfilled, the procedure converges to the fixed point of the map (\ref{extreme}), i.e. to the maximum-likelihood estimate of photon statistics \cite{Dempster1977,Vardi1993}.

\begin{figure*}[h!]
	\centering
	\includegraphics[width=0.6\textwidth]{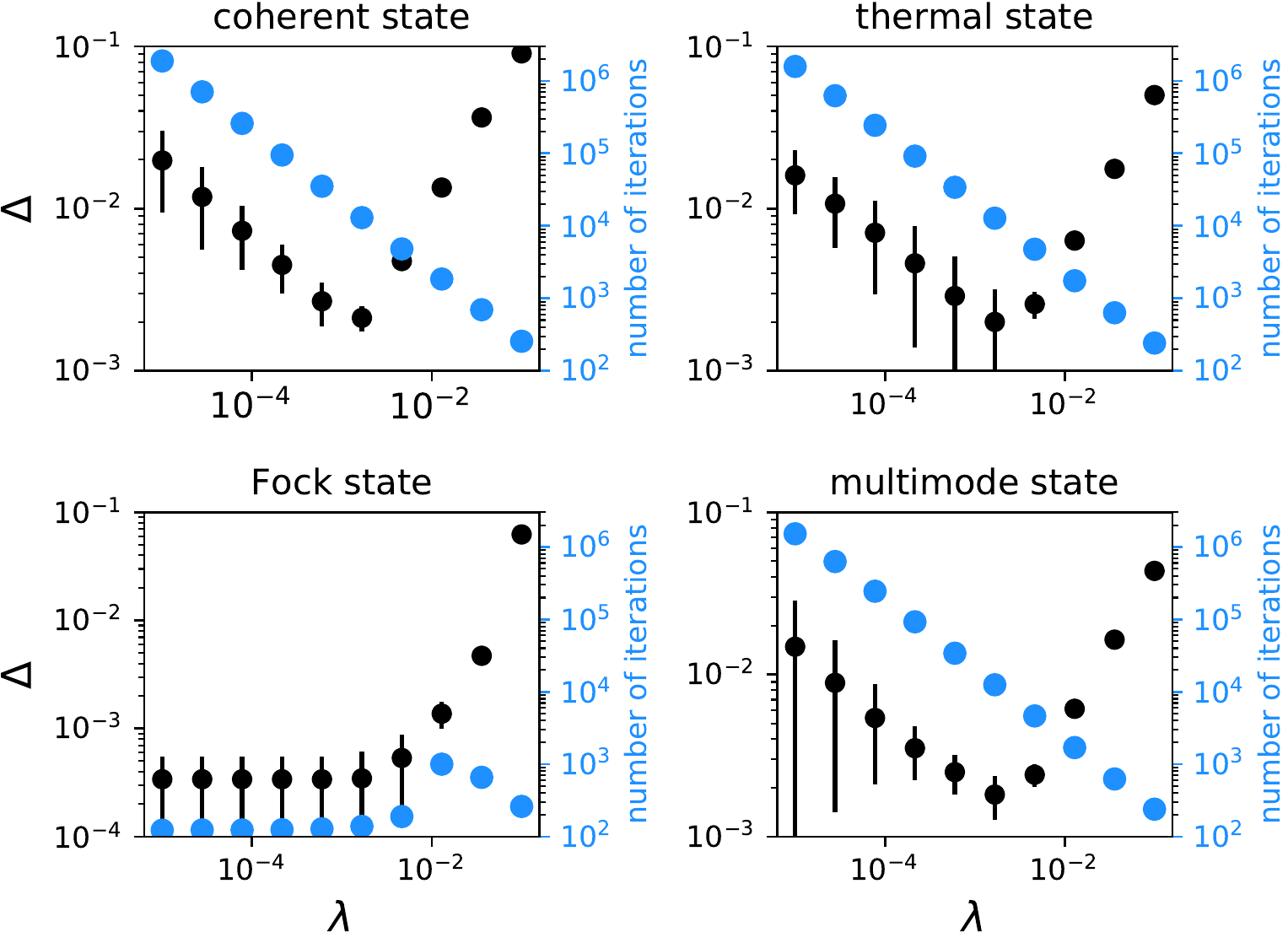}
	\caption{Accuracy and speed of photon statistics retrieval using EME algorithm versus the strength of entropy regularization characterized by the parameter $\lambda$. The total variational distance $\Delta$ characterizing the accuracy (black) and the number of EME iterations required (blue) are plotted for four different photon statistics. Shown are: coherent state with $\left \langle n \right \rangle=5$, thermal state with $\left \langle n \right \rangle=2$, single-photon emitter ($N_{p}=1$), and multimode thermal states with $M_{th}=2$. The data are simulated numerically (from true photon statistics) using Monte Carlo approach. For each value of $\lambda$ the statistics retrieval is performed several times for different data sets to evaluate the repeatability, which is represented by error bars of $\Delta$.}
	\label{lambda}
\end{figure*}

\textbf{Expectation-maximization-entropy algorithm.}
For underdetermined problems, when $n_\textrm{\scriptsize max}>M$, the EM algorithm will converge to a particular solution depending on the initial distribution $\{p_n^{(0)}\}$. All the possible solutions reaches the same value of the likelihood (given the data ${d_n}$) and cannot be distinguished by ML principle itself. 
In such cases, the common strategy is to allow for some kind of regularization or damping to select the most ``simple'' solution from the plateau of all ML solutions or, in other words, to prevent overfitting of the data. Entropy characterizes the solution complexity and its maximization reflects minimum prior information. Entropy maximization is frequently used for regularization of inverse problems in various applications like image reconstruction, seismology, and electromagnetic theory \cite{Smith1985,Parker1994}, and also in machine learning and quantum state estimation \cite{Bengio2005,Teo2011}.
Adopting this idea, we have applied entropy maximization to EM algorithm to obtain the most-likely estimate of photon statistics with the largest entropy. The resulting strategy not only offers improved fidelity of the retrieved statistics but also makes the iteration process faster.
The derivation of the expectation-maximization-entropy (EME) algorithm is analogous to derivation (\ref{loglikelihood})-(\ref{EM}) but the regularized functional $E$ is used instead of simple log-likelihood,
\begin{equation}\label{likent}
	E\left[\{p_n\}\right] = \sum_{m=0}^M d_m \ln c_m + \lambda \sum_{n=0}^{n_\textrm{\scriptsize max}} p_n \ln p_n
	- D \left( \sum_{n=0}^{n_\textrm{\scriptsize max}} p_n - 1 \right).
\end{equation}
Parameter $\lambda$ scales the entropy regularization relative to the likelihood maximization.
Performing variation of the log-likelihood-entropy functional $E$, eliminating the Lagrange multiplier $D$, and rewriting the extremal equation in the iterative form lead us to the EME algorithm
\begin{equation}\label{EME}
	\Pi_{n}^{(k)} p_n^{(k)} - \lambda \left( \ln p_{n}^{(k)} - S^{(k)} \right) p_{n}^{(k)} = p_n^{(k+1)}, \qquad \Pi_{n}^{(k)} = \sum_{m=0}^M \frac{d_m}{\left( \sum_{j} C_{mj} p_j^{(k)} \right)} C_{mn},
	\qquad S^{(k)} = \sum_{n=0}^{n_\textrm{\scriptsize max}} p_n^{(k)} \ln p_n^{(k)}.
\end{equation}
The initial iteration is chosen to contain no prior information about the statistics, $p_n^{(0)} = 1/(n_\textrm{max}+1)$, and the process is terminated based on the distance (\ref{stop}).
An implementation of this algorithm in Python is presented below.
We have performed hundreds of photon statistics retrievals using measured data and thousands retrievals based on Monte Carlo simulated data with not a single failure of the EME algorithm convergence. We have also verified that the retrieved photon statistics does not depend on the initial iteration.

Furthermore, we have performed a detailed analysis of accuracy and convergence speed as a functions of the regularization parameter $\lambda$ for various photon statistics including strongly non-classical sub-Poissonian states. 
In case of small values of $\lambda$ the EME approaches the common EM algorithm and the accuracy and repeatability of the solution decrease. For large $\lambda$ the entropy regularization prevails and the solution is less likely to reproduce the data - the accuracy drops. The optimum value of the regularization parameter is found to be $\lambda = 10^{-3}$ for all our data sets.
The same value is used throughout this work for all the performed photon statistics retrievals (for all tested sources).

\begin{figure*}[t]
	\centering
	\includegraphics[width=0.8\textwidth]{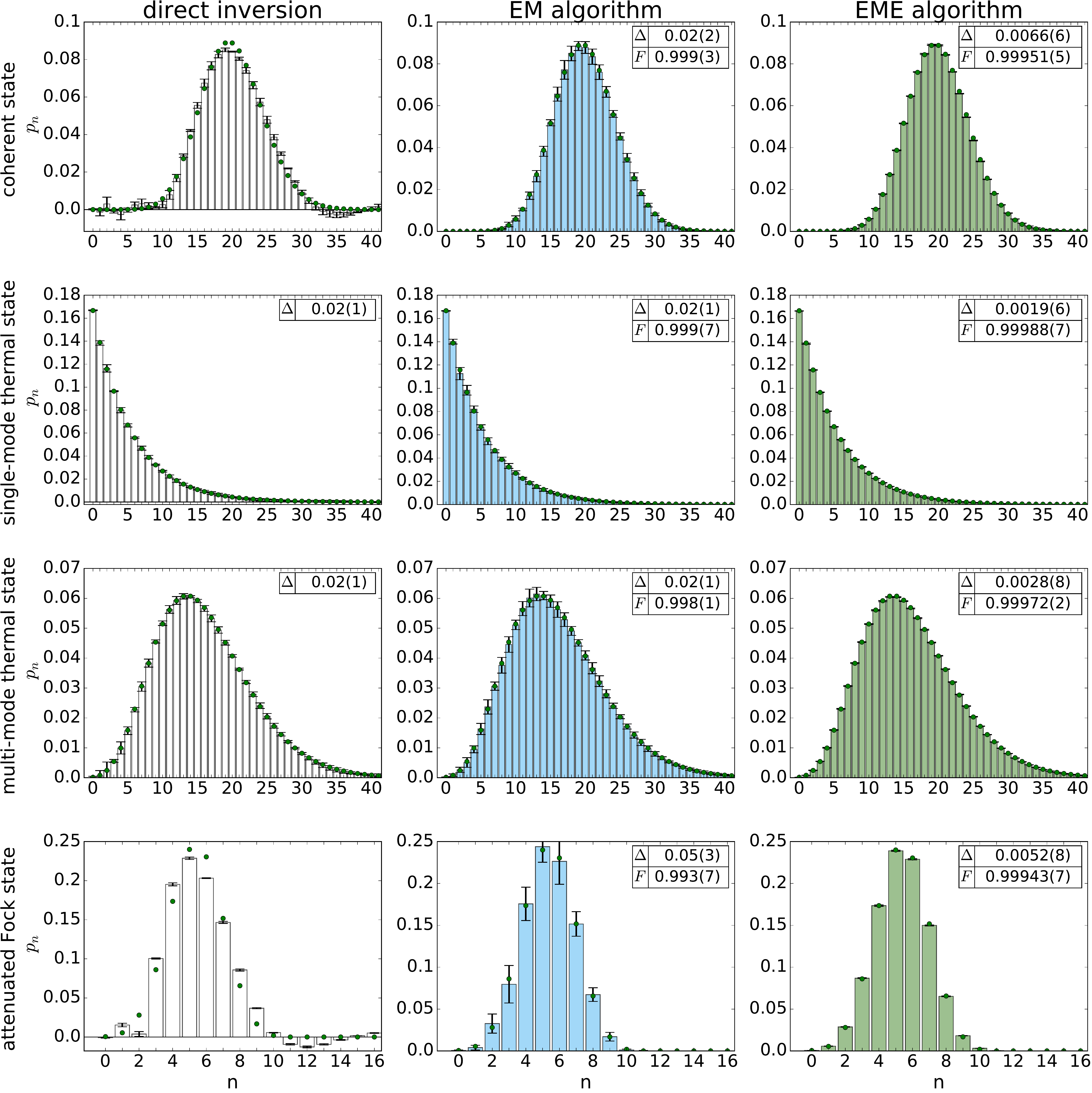}
	\caption{Photon statistics retrieval of various different states of light from numerically simulated click statistics. Each column shows a different retrieval method: direct inversion (Moore-Penrose pseudoinverse), EM algorithm, and EME algorithm, from left to right. The same amount of data (number of measurement runs) is used for each retrieval method to facilitate the comparison.
		The green points represent the corresponding true photon statistics.}
	\label{fig_retrieval_methods}
\end{figure*}

\newpage

\section*{Example \emph{EME} code}

\begin{Verbatim}[tabsize=4]
import math
import scipy.special as sc
import numpy as np
from scipy.misc import factorial

#Click statistics
c=np.array([6.73794700e-03,4.37104954e-02,1.27601677e-01,2.20741125e-01\
,2.50599060e-01,1.95082729e-01,1.05461930e-01,3.90945126e-02\
,9.51054071e-03,1.37104223e-03,8.89424261e-05])

def DET(mMax,nMax,eta):
det=np.zeros((mMax+1,nMax+1))
for m in range(mMax+1):
for n in range(nMax+1):
if m>n:
det[m][n]=0
elif m<n:
summary=[]
for j in range(0,m+1):
summary.append(((-1)**j)*sc.binom(m,j)*((1-eta)+((m-j)*eta)/mMax)**n)
det[m][n]=sc.binom(mMax,m)*np.sum(summary)
else:
det[m][n]=(eta/mMax)**n*(factorial(mMax)/factorial(mMax-n))	
return det

def EME(mMax,nMax,eta,det,l,c):
iterations = 10**10
epsilon = 10**(-12)
for j in range(0,len(c)):
pn=np.array([1./(nMax+1)]*(nMax+1))
iteration=0
while (iteration<iterations):
EM=np.dot(c/np.dot(det,pn),det)*pn
E=l*(np.log(pn)-np.sum(pn*np.log(pn)))*pn
E[np.isnan(E)]=0.0
EME=EM-E
dist = np.sqrt(np.sum((EME-pn)**2))
if dist<=epsilon:
break
else:
pn=EME
iteration+=1
return EME

mMax=10
nMax=50
eta=0.5
l=10**(-3)

det=DET(mMax,nMax,eta)
p=EME(mMax,nMax,eta,det,l,c)

print p
\end{Verbatim}

\section{Comparison of the retrieval algorithms}

We performed a numerical analysis comparing EME to other photon statistics retrieval methods (Fig.~\ref{fig_retrieval_methods}). We numerically simulated click statistics (using $M=10$ and $\eta=0.5$) of several known initial states and then applied direct inversion, EM algorithm and EME algorithm. To quantify the match between the real and estimated photon statistics, we used total variation distance and fidelity. The direct inversion method proved to be unsatisfactory, because non-negativity of the result is not guaranteed and therefore, some results do not represent a valid photon statistics. Those that do, exhibit the distance $2\times10^{-2}$, which is close to the results of the EM method. The EM algorithm guarantees positive-semidefinite results with average fidelity $\overline F = 0.997$. The total variation distances are similar to those obtained by direct inversion. An average distance $3\times10^{-2}$ is reached for all tested sources. Finally, the presented EME method gives the best match while always maintaining non-negativity. The average fidelity $\overline F = 0.9996$ and average distance is $4 \times 10^{-3}$. Particularly, the total variation distance of this method is smaller by an order of magnitude across all states. The EME therefore significantly improves the results for all kinds of simulated statistics.

\begin{figure*}[h!]
	\centering
	\includegraphics[width=0.7\textwidth]{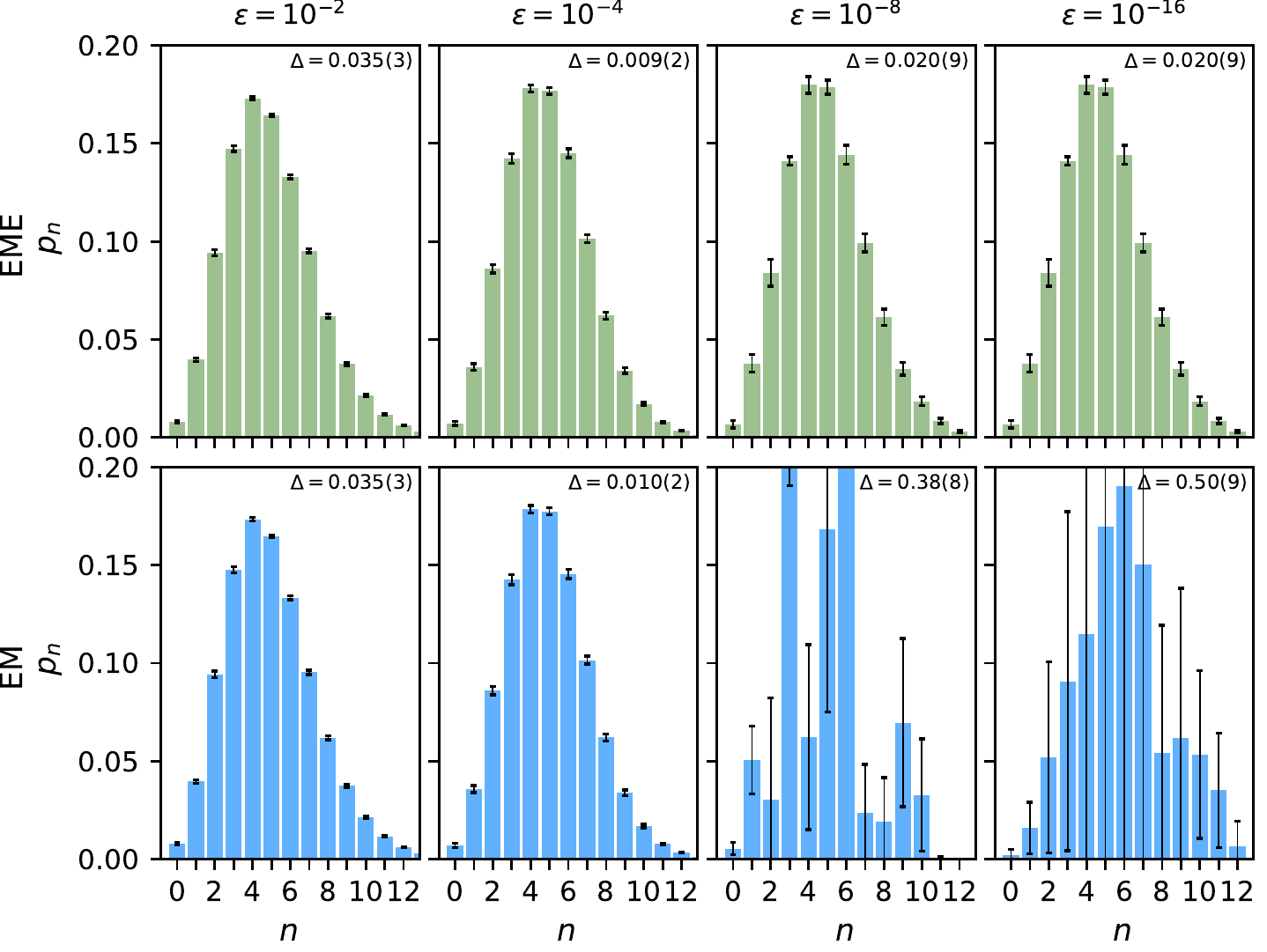}
	\caption{Poisson statistics $\langle n \rangle=4.95$ reconstructed by EME (top row) and EM (bottom row) as a function of the iteration cut-off distance $\epsilon$. All distributions are obtained from the same collection of 10 data sets, each containing $10^5$ measurement runs. Inset number (top right) denote total variation distances to the ideal Poisson distribution.}
	\label{fig_overfitting}
\end{figure*}

EME and EM exhibit significantly different convergence behavior with respect to the iteration cut-off distance $\epsilon$. For both methods, the inter-step distance decreases with the number of steps. For EM, the total variation distance to the expected photon statistics is non-monotone and eventually starts to rise. The result is that for low $\epsilon$ the retrieved photon statistics reveals considerable artifacts. EME does not show this issue. 
We demonstrate this effect on measured data for a Poissonian signal, see Fig.~\ref{fig_overfitting}. For both methods, the reconstructed photon statistics yield the same click statistics on the PNRD, but the ill-posed nature of the problem results in overfitting in the case of EM. For EME, the weak regularization eliminates this issue. We observed that this behavior is stronger for smaller data sets. It may seem that using a certain optimal value of $\epsilon$ would solve the issue. Unfortunately, the optimal value depends on the photon statistics. When measuring an unknown distribution, the value of $\epsilon$ cannot be set beforehand, because no expected distribution is available.

\begin{figure*}[h!]
	\centering
	\includegraphics[width=0.55\textwidth]{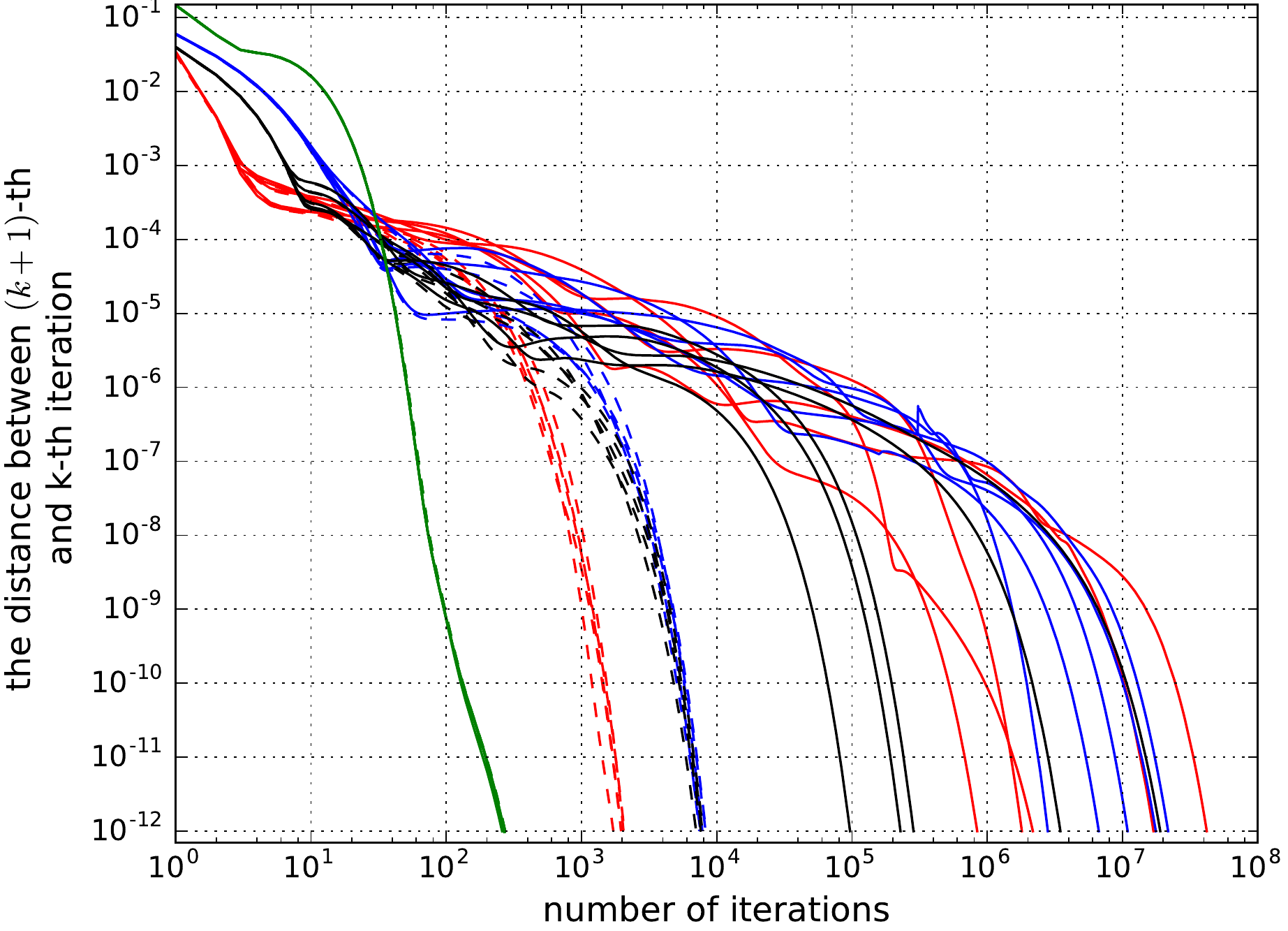}
	\caption{Photon statistics retrieval convergence demonstration of selected states: coherent state (blue), thermal state (red), 2-photon subtracted thermal state (black), and single-photon emitter (green). Shown are results for EM iterative process (full lines) based on Eq. (\ref{EM}) and for EME algorithm (dashed lines) based on Eq. (\ref{EME}). For each state five runs of individual retrievals were done.}
	\label{fig_convergence}
\end{figure*}

The convergence speed analysis of the retrieval methods with respect to the measured data is shown in Fig.~\ref{fig_convergence}. We compare the convergence of EM and EME in the case of a coherent state, a thermal state, a two photon-subtracted thermal state and a single-photon emitter.
Fig.~\ref{fig_convergence} shows that EME converges faster by orders of magnitude than EM. Only for a single-photon emitter, both methods are on par (the green lines overlap). While EM usually requires at least $10^5$ iterations, EME can do with less than $10^4$.

\section{Loss budget, multiple beam generation, and on-chip integration}

The presented measurements have been performed with PNRD detector not optimized for detection efficiency. The overall system efficiency is estimated to be $49(1)\%$. The efficiency parameter $\eta$ incorporated in photon statistics retrieval is chosen to be $0.5$ to assure that the efficiency of the PNRD model is the same or higher than of the actual PNRD detector used. The same value of the efficiency is used throughout this work for all the performed photon statistics retrievals (for all tested sources).

The non-unity system efficiency is caused by a sequence of five half wave plates and polarizing beam splitters with the total transmittance of $0.97^5$, two lenses and two fiber couplings with the transmittance of $0.88$, and the efficiency of SPAD detectors ranging from $0.6$ to $0.7$ with average value of $0.65$. Hence, $0.97^5 \times 0.88 \times 0.65 = 0.49$.
The efficiency can be improved by employing low-loss optics (especially polarizing beam splitters), anti-reflection coated fibers (transmittance 99\%), and super-conducting nanowire single-photon detectors (system efficiency 90\%). The improved efficiency can reach $0.985^5 \times 0.96 \times 0.9 = 0.8$. Based on the performed numerical simulations we expect that the resulting retrieved photon statistics will be nearly identical to the ones retrieved using the current version of the PNRD detector. The high-efficiency detector would find its application mainly in the case of low number of measurement runs and as heralding detector for a preparation of highly-nonclassical quantum states.

The efficiency can be improved significantly by a coherent detection such as homodyning with close-to-unity quantum efficiency photodiodes \cite{Lvovsky2009}. Also, homodyne detection provides full information about quantum state of light, including phase information. Unfortunately, the homodyning requires a proper (frequency adjusted) local oscillator, which is not accessible in many applications and for many sources like solid-state emitters, biomedical samples, and generally all multi-mode sources.

The discrete optical network employed in the reported PNRD features full reconfigurability and continuous tunability of splitting ratios, but extends over dozens square decimeters and limits the overall efficiency of the PNRD.
There are other ways of producing multiple beams of uniform intensity: diffraction gratings (diffractive beam splitter) \cite{Golub2004,Hermerschmidt2007}, multiple-beam plate splitters \cite{Webb1995}, and $M\times M$ fiber splitters and fan-outs. These solutions possess limited efficiency and no tunability and reconfigurability. On-chip integration offers a significant reduction in size \cite{Szameit2016}, however, the limited transmittance of a waveguide network and input/output coupling losses represent an issue. The tunability can be reached using interferometer networks with adjustable phase shifters \cite{OBrien2009onchip,Walmsley2009,Osellame2015,Laing2015}.

\section*{Acknowledgments}
We acknowledge the support from the Czech Science Foundation under the project 17-26143S.
This work has received national funding from the MEYS and the funding from European Union's Horizon 2020 (2014-2020) research and innovation framework programme under grant agreement No 731473 (project 8C18002). Project HYPER-U-P-S has received funding from the QuantERA ERA-NET Cofund in Quantum Technologies implemented within the European Union's Horizon 2020 Programme. We also acknowledge the support from Palack\'y University (projects IGA-PrF-2018-010 and IGA-PrF-2019-010).


%

\end{document}